\RequirePackage{amsmath}
\documentclass[runningheads]{llncs}
\usepackage{lipsum}

\newcommand\blfootnote[1]{%
  \begingroup
  \renewcommand\thefootnote{}\footnote{#1}%
  \addtocounter{footnote}{-1}%
  \endgroup
}

\usepackage{tabularx}
\usepackage{graphicx}

\begin{document}
\title{A Review of Homomorphic Encryption Libraries for Secure Computation}

%
%
\author{Sai Sri Sathya\inst{1}\thanks{Corresponding author 1, e-mail: origin@s20.ai } \and
Praneeth Vepakomma\inst{2,3}\thanks{Corresponding author 2, e-mail: vepakom@mit.edu } \and
Ramesh Raskar \inst{2,3} \and
 Ranjan Ramachandra\inst{1}\and
 Santanu Bhattacharya\inst{3}\thanks{The author is a collaborator with the Camera Culture Group, MIT Media Lab}}
\authorrunning{A Review of Homomorphic Encryption Libraries for Secure Computation}
%
\institute{S20.ai
\\
 \and
Massachusetts Institute of Technology\\ 
\and 
Camera Culture Group, MIT Media Lab
}
\maketitle              
\blfootnote{The authors would like to thank Niyaatii Swami for proof reading the paper}
\vspace{-0.6cm}
\begin{abstract}
In this paper we provide a survey of various libraries for homomorphic encryption. We describe key features and trade-offs that should be considered while choosing the right approach for secure computation. We then present a comparison of six commonly available Homomorphic Encryption libraries - SEAL, HElib, TFHE, Paillier, ELGamal and RSA across these identified features. Support for different languages and real-life applications are also elucidated.  

\end{abstract}
\section{Introduction}
Homomorphic Encryption is a method of secure computation on encrypted data (ciphertext) such that the result of the computation is also a ciphertext. Once this resultant ciphertext is decrypted, the decrypted result should match the output of operations on the corresponding unencrypted (plaintext) data. 

For example, a hospital that has a significant amount of private and sensitive information on patients can homomorphically encrypt the data and send it to a third party for analysis. The third party can perform calculations on encrypted data and send the results (also encrypted) back to the hospital. The hospital can then view the results by decrypting the data using a private key.  

There are several schemes of homomorphic encryption categorized based on the number of operations allowed on the encrypted data. For a cryptosystem to be \textbf{Fully Homomorphic} (FHE) it should support any number of arbitrary computations. Brakerski-Gentry-Vaikuntanathan (BGV) \cite{FHEwoBootstrap} and CGGI \cite{FasterPackedFHEBootstrapping,FasterFHE} are examples of Fully Homomorphic schemes. In practice, Fully Homomorphic schemes have tremendous overhead and are computationally very expensive. \textbf{Somewhat Homomorphic Encryption} (SWHE) schemes are practically more feasible but allow only certain operations on encrypted data and limit the number of computations as the Ciphertext size increases with each step due to noise. Some examples of Somewhat Homomorphic Encryption schemes are  \cite{SomewhatFHE,YaoProtocols,NCSUP1,DNFCipher,branchingEnc}. \textbf{Partially Homomorphic Encryption} (PHE) schemes such as \cite{verifiableBallot,DenseProbEnc,RivestDigSig,ProbEncPoker,Elgamal,ResiduesPKC,factoringPKC,Pailier,PailierAppl,LatticeCrypt,PailierElliptic} allow only one type of operation any number of times - either addition or multiplication on encrypted data as compared to Somewhat Homomorphic schemes that support both. Generation of such schemes continues to be an active area of research and development of standards for homomorphic encryption has recently taken-off as described in \cite{SecurityHE,HES}. Below is an example to introduce the high-level concept of homomorphic encryption:

\begin{enumerate}
  \item Let m be the plaintext message
  \item Let a shared public key be a random odd integer $p$
  \item Choose a random large $q$, small $r$, $(|r| \leq p / 2)$
  \item Ciphertext $c = pq + 2r + m$ (Ciphertext $c$ is close to multiple of $p$)
  \item Perform homomorphic addition / multiplication as required
  \item Decrypt: $m = ( c \mod p ) \mod 2$
\end{enumerate}
In this case, the corresponding homomorphic operations of addition and multiplication are given below:\\\\
\textbf{Homomorphic Addition}
$$c_1 = q_1 * p + 2 * r_1 + m_1$$
$$c_2 = q_2 * p + 2 * r_2 + m_2$$
$$c_1 + c_2 = ( q_1 + q_2 ) * p + 2 * (r_1 + r_2) + (m_1 + m_2)$$ \\
\textbf{Homomorphic Multiplication}
$$c_1 = q_1 * p + 2 * r_1 + m_1$$
$$c_2 = q_2 * p + 2 * r_2 + m_2$$
$$c_1 * c_2 = ( (c_1 * q_2) + q_1 * c_2 * q_1 * q_2 ) * p  + 2 ( 2 * r_1 * r_2 + r_1 * m_2 + m_1*r_2 ) + m_1*m_2$$ 

If more complicated functions that require operations other than addition and multiplication need to be homomorphically encrypted, an alternative would be to generate a polynomial approximation (using Taylor series for example) and then apply homomorphic encryption on the resulting polynomial instead.  

Homomorphic encryption libraries are based on different schemes and hence feature different behavior. Microsoft's SEAL(V2.3.1) \cite{seal} is based on BFV \cite{SomewhatFHE}, HElib is based on BGV \cite{FHEwoBootstrap} and TFHE is based on CGGI \cite{FasterPackedFHEBootstrapping,FasterFHE}. 

\section{Features of Homomorphic Encryption Libraries 
}In this section we introduce important features of homomorphic encryption libraries. Features such as asymmetry, negative computations, noise budget, recrypt, ciphertext packing, bootstrapping \cite{FHEwoBootstrap,FasterPackedFHEBootstrapping} and relinearization are discussed in subsections 2.1 and 2.2. In  sub-section 2.3, operations (atomic) allowed by various libraries are discussed and supported languages are also mentioned for all the  libraries.
\subsection{Basic features}
\subsubsection{Asymmetry} 
All homomorphic encryption libraries in this study have been implemented in an asymmetric manner where they use a pair of keys for encryption and decryption of data. To be specific, keys used in asymmetric cryptography include a public key to encrypt the plaintext data which may be shared widely and a private key to decrypt the encrypted result. This is different from symmetric cryptographic systems that use a single key to encrypt the plaintext as well as decrypt the ciphertext.

\subsubsection{Serialization}
Certain homomorphic encryption libraries such as SEAL, HElib and TFHE provide custom APIs to serialize (and deserialize) keys and ciphertexts for local storage and retrieval. Libraries that don\textsc{\char13}t provide this feature require the developers to create their own implementation of serialization (which could be challenging with the complex data type) unless the ciphertext can be represented in a primitive type such as String or BigInteger using the same library.

\subsubsection{Negative computations} 
Negative computations correspond to subtracting operand 1 with operand 2 (where operand 2 $>$ operand 1). This means the result of this computation should be a negative number. Microsoft SEAL(V2.3.1) uses BFV and Cheon-Kim-Kim-Song (CKKS) for encryption. In this scheme the integers or real numbers correspond to polynomials in a specifically chosen ring \cite{AbstractAlgebra}. Different kinds of encoders such as Integer, Scalar, Fractional, PolyCRTBuilder can be used to convert the integers/reals in the input data into the corresponding coefficients in the polynomial space. In SEAL, if the ciphertext is encoded using ‘Integer Encoder’ or ‘Fractional Encoder’ then the negative computations are supported. On the other hand, if the ciphertext is composed and encrypted using a ‘PolyCRTBuilder’ then the resultant ciphertext after homomorphic subtraction will not be negative. This is due to the limitations in the chinese remainder theorem when dealing with absolute values.

\subsubsection{Encryption Parameters, Ciphertext Size and Memory Requirements}
Implementing homomorphic encryption through any library requires certain encryption parameters such as polynomial modulus, coefficient modulus, plain modulus, noise standard deviation, a random generator, etc to be initialized. Choice of these parameters can significantly affect the size of the ciphertext, RAM required, noise budget (refer section 2.2.1), speed, performance and security \cite{seal} of the encryption. Size of the ciphertext is usually large with the complex Ciphertext data type in libraries like SEAL and HElib and operations such as matrix rotation that use Galois keys (in SEAL) could result in huge RAM requirements. 

\newcolumntype{U}{>{\centering\arraybackslash}m{2cm}}
\newcolumntype{V}{>{\centering\arraybackslash}m{1.2cm}}
{\renewcommand{\arraystretch}{2}%
\begin{table}[h]
\resizebox{\textwidth}{!}{%
\begin{tabular}{|m{3cm}|V|V|U|U|U|U|}
\hline 
\centering\textbf{Basic Features}                     & \textbf{SEAL} & \textbf{HElib} & \textbf{TFHE} & \textbf{Paillier} & \textbf{ELGamal} & \textbf{RSA} \\ \hline
\textbf{Asymmetric}                             & Yes           & Yes             & Yes           & Yes               & Yes               & Yes          \\ \hline
\textbf{Serialization and Deserialization of keys and ciphertexts}        & Yes            & Yes             & Yes            & No                & No                & No           \\ \hline
\textbf{Negative computations support}                & Yes           & No              & No            & No                & No                & No           \\ \hline
\textbf{Ciphertext size (less than 1MB for 1 input)} & No            & No              & Yes           & Yes               & Yes               & Yes          \\ \hline
\textbf{Can run on less than 2GB RAM}            & No           & Yes              & Yes            & Yes                & Yes                & Yes           \\ \hline
\end{tabular}
}
\caption{Comparison of Homomorphic Encryption libraries based on basic features}
\label{my-label}
\end{table}}

\subsection{Advanced features}
\subsubsection{Noise budget}
A noise term is generally appended to a ciphertext in the encryption operation to guarantee the security of the cryptosystem. This term could be an integer (if the scheme is based on integers) or a polynomial (if the scheme is based on polynomials) with coefficients in \{−1,0,1\}. Size of the term depends on the security and correctness properties of each system (for instance, a polynomial is typically considered small if all its coefficients are small). Homomorphic operations increase the noise and beyond a threshold, the resultant ciphertext would become too corrupt to be decrypted. Noise budget (invariant) is defined as the total amount of noise that can be added until the decryption fails. Addition and subtraction have a very small impact on noise compared to multiplication and Partially Homomorphic Encryption schemes are not affected by noise. 

\subsubsection{Recryption}
Recryption is a technique to re-generate the noise budget of a ciphertext that was depleted by arbitrary computations. Recryption boosts bounded-depth homomorphism to unbounded-depth homomorphism. This implies that the noisy ciphertext can be converted into a noise-free ciphertext (of the same plaintext) without the secret key \cite{GentryLWE,LWEPacked}. Libraries that do not have recryption functionality implemented, provide no means of converting a noisy ciphertext to a noise-free ciphertext. They therefore limit the number of arbitrary computations on a ciphertext.
\subsubsection{Ciphertext Packing}
In some homomorphic encryption libraries such as SEAL and HElib, a list of plain values can be packed into a single ciphertext vector by a technique called ciphertext packing using the Chinese Remainder Theorem (CRT) \cite{LWEPacked}. Homomorphic operations are performed on these vectors, component-wise in a SIMD (Single Instruction Multiple Data) fashion. Ciphertext packing is used to achieve a nearly optimal homomorphic evaluation (upto polylogarithmic factors). Homomorphic operations act element-wise between encrypted matrices, allowing the user to obtain speed-ups of several orders of magnitude in naively vectorizable computations.
\subsubsection{Bootstrapping}
In certain homomorphic encryption schemes, arithmetic operations on ciphertext can be performed using basic gates (AND, OR, NOT, etc) but arbitrary operations could reduce the available noise budget. Bootstrapping \cite{FHEwoBootstrap,FasterPackedFHEBootstrapping} is a technique to remove noise by passing a ciphertext and encrypted private key into a circuit that represents the decryption algorithm of a FHE scheme. This results in a new ciphertext that corresponds to the original ciphertext but with no noise. In the TFHE library, after every gate-by-gate operation, bootstrapping is applied on the resultant ciphertext and hence any number of arbitrary operations can be performed. 
\subsubsection{Relinearization}
Two input ciphertexts of sizes m and n respectively result in a ciphertext of the size m+n-1 after multiplication. Consumption of the noise budget is also much higher during multiplication especially when the input ciphertexts sizes are huge. Relinearization reduces the size of the resultant ciphertext after a multiplication operation to the initial size. A ciphertext of size k + 1 when relinearized produces a ciphertext of size k. After repeated steps, this can result in a ciphertext of size 2 that can be decrypted using a smaller degree decryption function to yield the same result \cite{seal}. Thus, relinearization of resultant ciphertext after multiplication, can significantly improve the performance on the subsequent operations although relinearization by itself has both a computational cost and a noise budget cost.
\subsubsection{Multithreading}
In homomorphic encryption libraries, multithreading corresponds to APIs exposed by the libraries being thread safe. Thread-safe APIs help avoid deadlock and ease effective inter thread communication. Most of the tools in SEAL such as Encryptor, Decryptor, PolyCRTBuilder, and Evaluator are thread-safe by default. HElib can be multithreaded by setting NTL\_THREADS=on, -DFHE\_THREADs, -DFHE\_DCRT\_THREADS flags before making the project. In Partial Homomorphic Encryption libraries discussed in the paper, multithreading is not supported. 

{\renewcommand{\arraystretch}{2}%
\begin{table}[ht]
\resizebox{\textwidth}{!}{%
\begin{tabular}{|m{3cm}|V|V|U|U|U|U|}
\hline
\centering\textbf{Advanced Features}               & \textbf{SEAL} & \textbf{HElib} & \textbf{TFHE} & \textbf{Paillier} & \textbf{ELGamal} & \textbf{RSA} \\ \hline
\textbf{Noise affected after each computation} & Yes           & Yes             & Yes           & No                & No                & No           \\ \hline
\textbf{Recryption}                               & No            & Yes             & Yes           & N/A               & N/A               & N/A          \\ \hline
\textbf{Ciphertext \r packing}         & Yes           & Yes             & No            & No                & No                & No           \\ \hline
\textbf{Relinearization}                       & Yes           & Yes             & No            & N/A               & N/A               & N/A          \\ \hline
\textbf{Multithreading}                        & Yes           & Yes             & No            & No               & No               & No          \\ \hline
\end{tabular}%
}
\caption{Comparison of Homomorphic Encryption Libraries based on advanced features}
\label{my-label-2}
\end{table}}

\subsection{Operations}
\subsubsection{Ciphertext Comparison} Two ciphertexts can be compared for equality, greater than, greater than or equal to, less than or less than or equal to. TFHE allows evaluating an arbitrary boolean circuit composed of binary gates, over encrypted data. A custom comparator circuit can be used to perform comparisons using TFHE. In SEAL and HElib, a Binary-Encoder must be used to generate a ciphertext comprising of only 0s and 1s. Two such ciphertexts can then be compared in a bit-wise manner. This process is time-consuming and less secure. A computer with limited resources can potentially decrypt a ciphertext by randomly comparing it with a known ciphertext. Due to this security threat, HE libraries do not readily expose a comparison API. 
\subsubsection{Division}
BGV or BFV schemes do not allow division of ciphertexts due to the randomness and complexity of the ciphertext. It’s possible to approximate division but using all kinds of expansions. In fully homomorphic encryption, division of ciphertext $A$ and ciphertext $B$ is performed by computing the inverse of ciphertext $B$ (decrypt, inverse and encrypt) and multiplying the inverse ciphertext by ciphertext A (multiplicative inverse). Another technique is through recursive subtraction. Recursive subtraction can work only if $A \% B$ is equal to zero.
\subsubsection{Boolean Operations}
Some homomorphic encryption libraries that are based on Secure Multilayer Perceptron \cite{perceptronHE} and Doubly Permuted Homomorphic Encryption \cite{PPVLDPHE} allow evaluating an arbitrary boolean circuit composed of binary gates over encrypted data.
\subsubsection{Matrix Operations}SEAL exposes an API to perform matrix rotation and element-wise addition, multiplication and subtraction. In PHE libraries, first a ciphertext matrix has to be created by performing element-wise encryption on a nxn plaintext matrix and then custom logic has to be implemented to perform rows and column rotation.
\subsubsection{Exponentiation}
Exponentiation of ciphertexts is usually A raised to the power of $B (A ^ B)$ where $A$ and $B$ are ciphertexts. Current FHE libraries only provide an implementation to raise a ciphertext base with a plain text exponent. This is accomplished through repetitive multiplication of the ciphertext. Eg: $3 ^ 4$ is $3 * 3* 3 * 3 = 81$. The same can be accomplished on PHE schemes. In additive PHE scheme, $3 ^ 3$ can be calculated as $3 ^ 3 = 27   |   3 + 3 + 3   = 9   |   9 + 9 + 9 = 27$.
\subsubsection{Add Plain, Subtract Plain, Multiple Plain}
Homomorphic operations are usually carried out between two cipher texts. If one of the operands could be a plaintext it could significantly improve the performance. The size of the resultant ciphertext remains the same as the input ciphertext and the relinearlization step could be skipped. SEAL provides functions to perform addition, subtraction and multiplication of a ciphertext with a plaintext. The ‘plain’ operations are implemented in SEAL as Evaluator::add\_plain, Evaluator::sub\_plain and Evaluator::multiply\_plain.

{\renewcommand{\arraystretch}{2}%
\begin{table}[h!]
\resizebox{\textwidth}{!}{%
\begin{tabular}{|m{3cm}|V|V|U|U|U|U|}
\hline
\centering\textbf{Operations}                                      & \textbf{SEAL} & \textbf{HElib} & \textbf{TFHE}               & \textbf{Paillier}           & \textbf{ELGamal}                 & \textbf{RSA}                      \\ \hline
\textbf{Addition, \r Subtraction}                                                & Yes           & Yes             & Yes                         & Yes                         & No                                & No                                \\ \hline
\textbf{Multiplication}                                          & Yes           & Yes             & Yes                         & No                          & Yes                               & Yes                               \\ \hline
\textbf{Comparison}                                              & No            & No              & No                         & No                          & No                                & No                                \\ \hline
\textbf{Division}                                                & No            & No              & No                          & No                          & No                                & No                                \\ \hline
\textbf{Boolean \r operations} & No            & No              & Yes                         & No                          & No                                & No                                \\ \hline
\textbf{Bitwise operations}                                      & Yes           & Yes             & Yes                         & No                          & No                                & No                                \\ \hline
\textbf{Matrix operations}                                       & Yes           & Yes             & No                          & No      & No      &No       \\ \hline
\textbf{Exponentiation}                                          & Yes           & Yes             & No & No & No & No \\ \hline
\textbf{Square}                                                  & Yes           & Yes             & Yes                         & No                          & Yes                               & Yes                               \\ \hline
\textbf{Negation}                                                & Yes           & Yes             & No                          & No                          & No                                & No                                \\ \hline
\textbf{Add Plain, Subtract Plain, Multiply Plain}                                          & Yes           & No              & No                          & No                          & No                                & No                                \\ \hline
\end{tabular}%
}
\caption{Different operations supported by Homomorphic Encryption libraries}
\label{my-label-3}
\end{table}}

{\renewcommand{\arraystretch}{2}%
\begin{table}[h]
\resizebox{\textwidth}{!}{%
\begin{tabular}{|m{3cm}|V|V|U|U|U|U|}
\hline
\centering\textbf{Languages} & \textbf{SEAL} & \textbf{HElib} & \textbf{TFHE} & \textbf{Paillier} & \textbf{ELGamal} & \textbf{RSA} \\ \hline

\textbf{C++}               & Yes           & Yes             & No            & Yes               & Yes               & Yes          \\ \hline
\textbf{Python}            & Yes           & Yes             & No            & Yes               & Yes               & Yes          \\ \hline
\textbf{Java}              & No            & No              & No            & Yes               & Yes               & Yes          \\ \hline
\textbf{C}                 & No            & No              & Yes           & No                & No                & No           \\ \hline
\end{tabular}%
}
\caption{Homomorphic Library implementations across programming languages}
\label{my-label-4}
\end{table}}

%
%
%
%
\section{Applications}
The need to create models or derive predictions from confidential distributed datasets is a commonly surfacing theme in many industries. For example, medical information might be distributed across multiple clinics.  \cite{ApplHE} outlines various potential real-world applications of homomorphic Encryption. Some of the emerging applications are:  
\subsection{Healthcare}
In healthcare, maintaining privacy of patients’ information is critical and therefore, their private date is often protected by law. However, sharing and computing on information that is distributed across systems is important for diverse use cases such as coordinated patient care, fraud billing and reimbursements. It is therefore difficult to strike a balance between risk and utility. For example, in 2018, there were 11 large HIPAA enforcement actions with an average fine of \$1.9 million \cite{ComplianceFines}. Homomorphic Encryption can help balance the risk  vis-a-vis utility by enabling the analysis of billing records across patient data to uncover potential cases of fraud reimbursement or billing, without violating the patient’s privacy.
\subsection{Financial Services}
Clients and businesses in the financial services work with confidential information. Consequently, data, the models and functions computed on them are often considered proprietary and confidential. Data in financial services functions may even be a continuous stream reflecting the most up-to-date information necessary for decisions making and is often a result of exclusive research or data feeds available to a particular client and is often, very expensive. Homomorphic Encryption provides the appropriate way to evaluate and run both these data and functions privately.  For instance, a client can upload an encrypted version of the function to the cloud, and the streaming data, on which the functions / models run could be encrypted using the customer's public key and uploaded to the cloud. 
\subsection{Smart Grid}
Consider a smart grid consisting of multiple microgrids such as solar panel generators used by individuals. Each node in such a grid generates useful data like electrical generation and usage, temperatures of physical equipments, energy flows, etc. In case of a generator, if the nodes belong to a smart grid, then measurements include current energy usage, smart lights, sensors in use etc. 

When the municipality or any other government entity wants an aggregate measure, or an alert about the data, they can use Homomorphic Encryption for computing data from nodes. They can do this without violating the terms of business contracts that prohibit them from disclosing confidential information such as usage of energy in a particular mall, or the location of the surveillance cameras in a household. In this way, they can develop trust and improve credibility of the smart grids with the public. Homomorphic encryption plays an important role in achieving this.
\subsection{Genomics}
Private data generated from sequencing human genome for complex disease or epidemiology can be a powerful tool in developing a cure or a therapy/treatment for the disease. DNA and RNA sequences can be generated rapidly and consequently, large amount of such sequences are now available in laboratories and medical institutes. However, significant challenges exist in sharing this data. \cite{PGAHE}. Individual DNA sequences are as unique as fingerprints - they can be tracked down to an individual and can determine say for e.g.  if they are susceptible to Alzheimer's disease or heart attack. Existing rules for protecting genomics data has created a lot of limitations for the researchers. Homomorphic Encryption can enable researchers to speed up sharing information while safeguarding privacy of the individuals and thus significantly speed up discovery.  
\section{Conclusion} In this paper we survey and compare libraries across various dimensions for homomorphic encryption. These techniques enable us to perform computations on encrypted data as against having to decrypt data in order to perform computations. In this way, it allows for collaborative computing between multiple parties via encrypted ciphertexts.  Although the field is rapidly progressing on the theoretical front, there has been significant recent progress in making it practical from an application/practical standpoint. Both these factors are crucial for rapid adoption and further development of this field. 

Applications of homomorphic encryption primarily involve distributed applications in diverse sectors such as healthcare, smart grids or genomics.  In these applications, ciphertexts, public keys, and other low-level information needs to be shared between data providers, encrypted computing hosts, and the desired recipients of the results of the computation. 

There are many scenarios, such as the one mentioned in healthcare detection or genomics research, where these applications are currently almost impossible to develop due to technical or legal reasons. In cases where the technology is available, one still has to cross the expensive and time-consuming barrier of legal processes, driven by the need of maintaining strict privacy. We can however hope, that practical homomorphic encryption would lead to a dramatic rise in applications in cloud and edge computation where privacy is critical.  Our intent is to share our learnings, motivate our colleagues and help the progress of the research and technology community. 
\section{Other Homomorphic Encryption Libraries}
 \cite{AwesomeHELibs} publishes an exhaustive list of Homomorphic Encryption Libraries:\\
HEAAN - Scheme with native support for fixed point approximate arithmetic\\
FHEW - Homomorphic Encryption library based on Fast Fourier Transform\\
Λ ○ λ - Haskell library for ring-based \cite{AbstractAlgebra} lattice cryptography that supports FHE\\
NFLlib - NTT-based Fast Lattice library\\
PALISADE - Lattice encryption library\\
Pyfhel - PYthon For HElib\\
libshe - Symmetric SWHE library based on DGHV scheme\\
cuHE - GPU-accelerated HE library for NVIDIA CUDA-Enabled GPUs\\
cuYASHE - Based on leveled FHE scheme YASHE for GPGPUs\\
python-paillier - PHE based on Paillier scheme\\
krypto - C++ implementation of multivariate quadratic FHE\\
petlib - Python library that implements a number of Privacy Enhancing Technologies\\


\end{document}